\documentclass[12pt,epsfig]{article}
\usepackage{graphicx}
\def\beq{\begin{equation}}
\def\eeq{\end{equation}}
\def\bea{\begin{eqnarray}}
\def\eea{\end{eqnarray}}
\newcommand{\sech} { {\rm sech}}

\begin{document}

\begin{center}
{\Large \bf \sf Reciprocity in parity violating non-Hermitian systems
  }

\vspace{1.3cm}

{\sf Ananya Ghatak \footnote{e-mail address: gananya04@gmail.com}
Brijesh Kumar Mourya\footnote{e-mail address: brijeshkumarbhu@gmail.com}
Raka Dona Ray Mandal \footnote{e-mail address: rakad.ray@gmail.com}
and Bhabani Prasad Mandal \footnote{e-mail address:
bhabani.mandal@gmail.com,\ \ bhabani@bhu.ac.in  }}

\bigskip

{
Department of Physics,
Banaras Hindu University,
Varanasi-221005, INDIA. \\ }

\bigskip
\bigskip

\noindent {\bf Abstract}

\end{center}

Reciprocity is shown so far only when the scattering potential is either real or parity symmetric complex.
We extend this result for parity violating complex potential by considering several explicit examples: (i)
we show reciprocity for a PT symmetric (hence parity violating) complex potential which admits penetrating 
state solutions analytically for all possible values of incidence energy and (ii) reciprocity is 
shown to hold at certain discrete energies for two other parity violating complex potentials.

\medskip
\vspace{1in}

\newpage
\section{Introduction} 
Over a decade and half fully consistent quantum theory for non-Hermitian systems \cite{ben4}-\cite{benr} 
have been developed extensively with its application in different branches of physics \cite{opt1}-\cite{cal}. Recently scattering due to complex potentials has attracted huge attention due to its rich 
properties and wide applicabilities and usefulness in the study of different optical 
systems \cite{ opt1}-\cite{eqv1}. Exceptional points (EPs) \cite{ep0}-\cite{ep2}, spectral 
singularity (SS) \cite{ss1}-\cite{ss3}, invisibility  \cite{aop}-\cite{inv1}, reciprocity  
\cite{aop}-\cite{resc}, critical coupling (CC) \cite{cc0}-\cite{cc4} and coherent perfect absorption (CPA) \cite{cpa00}-\cite{cpa4} are among the interesting features of complex 
scattering. In particular, the CPA which is time reversed counter part of lasing effect has 
become the center of all such studies in optics due to the discovery of 
anti-laser \cite{cpa00}-\cite{cpa011} which has a number of technological implications. Other exciting feature is 
reciprocity which is the topic of discussion in the present work. In the case of scattering due 
to real potential, transmission coefficient ($R_l$) for left incident particle is always equal 
to that ($R_r$) of for the right incident particle, and the system is called reciprocal.
Reciprocity holds good even for complex potentials which respect parity \cite{reci1, reci2}. 
However reciprocity is known to be violated in the case of complex potentials which are not parity invariant.
In fact  PT symmetric non-Hermitian systems are always parity violating and hence non-reciprocal. Reciprocity has not been investigated in details for the PT symmetric
non-Hermitian systems and to the best of our knowledge not a single example of parity 
violating PT-symmetric reciprocal system is known till date. Therefore it is worth investigating reciprocity in parity violating non-Hermitian systems.
In this work
 we show that reciprocity can be restored even for the parity violating non-Hermitian systems. 
We consider three specific examples in support of our claim. In the first example we consider 
complexified Morse type potential \cite{aop} which admits bound, reflecting, penetrating and free state
solutions depending on the energy of the incident particle. We show that reciprocity is restored for the entire 
range of incident energy even though the potential is parity violating and complex, admitting 
penetrating state solutions. In the case of 
reflecting states of the same model and in the case of parity violating (PT symmetric or non PT-symmetric) non-Hermitian double delta 
potential \cite{ram}, we show that reciprocity is valid only at some specific value of incident energy. These results
are shown graphically for the last two models. The most interesting  part of the non-Hermitian
double delta potential is that the reciprocity is restored only at extreme situations where 
spectral singularity or reflectionlessness occurs.

The plan of the paper is as follows. We systematically analyse the conditions of reciprocity 
in the case of both parity symmetric as well as parity violating real potential in section 2.
In section 3 we use the result of section 2 to show the reciprocity for three parity violating 
complex potentials. Section 4 is kept for summary and discussion.

\section{ Reciprocity : Real Potential} 
In this section we obtain the general condition for reciprocity in usual quantum mechanics.
The arguments are valid for both real and complex scattering and will be used in the later section.
To demonstrate this we consider the general solution of Schroedinger equation for the scattering state
as the superposition of the two independent solutions as,

\beq
\psi(x)=a U(x)+ b V(x)
\label{sc}
\eeq
where $U(x)$ and $V(x)$ are two independent scattering state solutions and $a, b$ are any complex numbers. Now in the case of scattering,
$U(x)$ and $V(x)$ are written in general asymptotic form, so that the wave function for $x\rightarrow +\infty$ is written as,
\bea
\psi^+(x) &=& a (u_1^+(k) e^{ikx}+u_2^+(k) e^{-ikx})+ b (v_1^+(k) e^{ikx}+v_2^+(k) e^{-ikx}) \nonumber \\
&=&\Big( au_1^+(k)+b v_1^+(k)\Big) e^{ikx}+\Big( au_2^+(k)+b v_2^+(k)\Big) e^{-ikx}
\label{+}
\eea
Similarly for $x\rightarrow -\infty$ the wave function is written as,
\beq
\psi^-(x) = \Big( au_1^-(k)+b v_1^-(k)\Big) e^{ikx}+\Big( au_2^-(k)+b v_2^-(k)\Big) e^{-ikx}
\label{-}
\eeq
From the Eqs. (\ref{+},\ref{-}) we can calculate the scattering amplitudes.

For left incidence:
 $au_2^+(k)+b v_2^+(k)=0$, and hence $r_l$ is evaluated as,
\beq
r_l=\frac{u_2^+ v_2^-(k)-v_2^+u_2^-(k)}{- v_2^+u_1^-(k)+u_2^+ v_1^-(k)}
\label{rl}
\eeq

For right incidence:
$au_1^-(k)+b v_1^-(k)=0$, so
\beq
r_r=\frac{v_1^-u_1^+(k)-u_1^- v_1^+(k)}{ -v_2^+u_1^-(k)+u_2^+ v_1^-(k)}
\label{rr}
\eeq
From Eqs. (\ref{rl},\ref{rr}) the condition for reciprocity ($R_l=R_r$)is written as,
\bea
u_2^+ v_2^-(k)-v_2^+u_2^-(k)&=& v_1^-u_1^+(k)-u_1^- v_1^+(k) \nonumber \\
\mbox{or} \ \ \mid u_2^+ v_2^-(k)-v_2^+u_2^-(k)\mid &=&\mid v_1^-u_1^+(k)-u_1^- v_1^+(k)\mid
\label{ler}
\eea
Now we discuss parity symmetric and parity violating cases separately using these
general results.

\subsection{ Parity symmetric potential}

For a parity symmetric potential (real or complex) the general wave functions are written in terms of the odd and even parity solutions. 
We consider  $U(x)=U(-x)$ and $V(x)=-V(-x)$. This implies
\beq
U(x\rightarrow \infty) \equiv U^+(x)\longrightarrow   U^+(-x) \equiv  U(x\rightarrow -\infty
 )=U^-(x) \ ; 
\label{Ps}
\eeq
and
\beq
V(x\rightarrow \infty) \equiv V^+(x)\longrightarrow   V^+(-x) \equiv  -V(x\rightarrow -\infty
 )=-V^-(x) \ ; 
\label{Ps}
\eeq
From the above equation Eq. (\ref{Ps}) we see that $U^+(-x)=U^-(x)$ and hence we obtain,
\bea
u_1^+(k) e^{-ikx}+u_2^+(k) e^{ikx}&=&u_1^-(k) e^{ikx}+u_2^-(k) e^{-ikx}, \nonumber \\
\mbox{i.e.} \ \ u_1^+(k)=u_2^-(k) \ &;& \ u_2^+(k)=u_1^-(k)
\label{u1u2}
\eea
Similarly due to $V^+(-x)=-V^-(x)$ we get,
\beq
v_1^+(k)=-v_2^-(k) \ ; \ v_2^+(k)=-v_1^-(k)
\label{v1v2}
\eeq
Due to these relations in Eq. (\ref{u1u2}) and Eq. (\ref{v1v2}), the condition of reciprocity
in Eq. (\ref{ler})is satisfied and the reflection coefficients as well as the amplitudes are
equal (i.e. $r_l=r_r$ and $R_l=R_r$) for arbitrary parity symmetric potential. \\

\subsection{Parity violating real potential} 
For a real potential, the wave function is always chosen as real in the form,
$\psi(x)=1/2\left\{\phi(x) +\phi^*(x) \right\}$ where $\phi^*(x)$ and $\phi(x)$ are two independent 
solution of Schroedinger equation. 
From the asymptotic forms of $\psi$ in Eq. (\ref{+}) and (\ref{-}) we have the following
equations for ${\psi^-}^*=\psi^-$ and by considering left incidence case we obtain
\bea
&& a^*{u_1^-}^*(k)+b^* {v_1^-}^*(k)= au_2^-(k)+b v_2^-(k) \nonumber \\
&\Rightarrow &\frac{-{v_2^+(k)}^*{u_1^-}^*(k)+{u_2^+(k)}^* {v_1^-}^*(k)}{{u_2^+(k)}^*}= \frac
{{-v_2^+}u_2^-(k)+{u_2^+(k)} v_2^-(k)}{u_2^+(k)} \nonumber \\
&\Rightarrow & N(r_l)=\frac{u_2^+(k)}{{u_2^+(k)}^*} \left [D(r_l)\right]^* \ , \ \ 
\mbox{thus} \ \ r_l=\frac{u_2^+(k)}{{u_2^+(k)}^*} \frac{\left [D(r_l)\right]^*}{\left [D(r_l)\right]}
\label{rl1}
\eea
where  $N(r_l)$ and $D(r_l)$ denote the numerator and denominator respectively in the expression of  $(r_l)$ in equation (4).
\beq
N(r_r)=\frac{u_1^-(k)}{{u_1^-(k)}^*} \left [D(r_r)\right]^* \ , \ \ \mbox{thus} \ \ r_r=\frac
{u_1^-(k)}{{u_1^-(k)}^*} \frac{\left [D(r_r)\right]^*}{\left [D(r_r)\right]}
\label{rr1}
\eeq
It is clear from equations (11) and (12) that the magnitudes of $r_l$ and $r_r$ are same, they only differ by a 
phase. Thus in this parity violating case $R_l= R_r$ even though $r_l\not =r_r$.
This proves the reciprocity for parity invariant real scattering case.

\section{Parity violating complex potential}

In this section we consider three different cases of parity violating complex potential to study 
the reciprocity. We complexify Morse type potential, which admits bound, reflecting, penetrating and free state
solutions in two different way. We show that the scattering state for this PT symmetric complex potential is reciprocal
 only at some specific incident energy, whereas the penetrating states are always reciprocal, even though 
the potential violate parity symmetry. In another case we consider PT-symmetric as well as
non PT-symmetric but Parity violating complex double delta potential to show that reciprocity 
is restored only at the energy values where SS and reflectionlessness occur.

\subsection{Morse-type potential with scattering states}

The one dimensional real Morse-type potential \cite{aop} is written as,
\beq
V^{I}(x)=V_0 \cosh ^2\mu' \{\tanh[(x-\mu' d)/d]+\tanh(\mu')\}^2 
\label{realm}
\eeq
This potential admits bound, reflecting, penetrating and free state solutions depending on 
the energy of the incident particle. We complexified this potential as $\mu'\rightarrow  i 
\mu $ so that the corresponding Hamiltonian becomes PT-symmetric non-Hermitian. The 
Schroedinger equation for this case is written as, 
\beq
d^2\psi_m /dz^2+[\epsilon-v \cosh (2 i\mu)-v \sinh (2 i \mu)\tanh z+v \cosh ^2(i\mu) \ \sech ^2 z]
\psi_m=0
\label{compm}
\eeq
where,
$z=(x-i\mu d)/d$, $v=(2md^2/\hbar^2)V_0$, $\epsilon=(2md^2/\hbar^2)E$. Here $z^{PT}=-z$ 
and $z^P\not=\pm z$, so the above potential is invariant only under PT-transformation.  The
scattering states solutions of Schroedinger 
equation for this non-Hermitian potential are,
\bea
U_{m} (z) &=& Ne^{\frac{1}{2}i(k_+-k_-)z}(e^z+e^{-z})^{\frac{1}{2}i(k_++k_-)z} \nonumber \\
&.& F\left(-\frac{1}{2}ik_+-\frac{1}{2}ik_-+\frac{1}{2}-\gamma,-\frac{1}{2}ik_+-\frac{1}{2}ik_-
+\frac{1}{2}+\gamma ;\mid 1-ik_+\mid;\frac{e^{-z}}{e^z+e^{-z}}\right)  \nonumber \\
V_{m} (z) &=& Ne^{\frac{1}{2}i(k_+-k_-)z}\Big( e^z+e^{-z}\Big)^{\frac{1}{2}i(k_++k_-)z}. {(\frac
{e^{-z}}{e^z+e^{-z}})}^{ik_+} \nonumber \\
&.& F\left(\frac{1}{2}ik_+-\frac{1}{2}ik_-+\frac{1}{2}-\gamma,\frac{1}{2}ik_+-\frac{1}{2}ik_-
+\frac{1}{2}+\gamma ;\mid 1+ik_+\mid;\frac{e^{-z}}{e^z+e^{-z}}\right) . \nonumber \\
\eea
where
\beq
\gamma =\sqrt{-v\cos^2 \mu +1/4} \ ; \ \ k_+=\sqrt{\epsilon -ve^{2i\mu}} ; \ \ k_-=\sqrt{\epsilon -ve^{-2i\mu}}
\label{k+-}
\eeq
The general scattering wave function for this Morse-like potential is written in the 
superposition form of the two independent solutions as,
\beq
\psi_m (z)=AU_{m}(z)+BV_{m}(z)
\label{m1}
\eeq
with $A$ and $B$ as the arbitrary constants using the standard properties of hyper-geometric functions the asymptotic forms of $\psi_1$
and $\psi_2$ are written in equivalent notations of Eqs. (\ref{+}) and (\ref{-}) as,
\bea
u_{m1}^+&=&N \ ; \ u_{m2}^+=0 \ ; \ v_{m1}^+=0 \ ; \ v_{m2}^+=N \ ; \ \nonumber \\
u_{m1}^-&=&N \ G2(k) \ ; \ u_{m2}^-=N \ G1(k) \ ; \ v_{m1}^-=N \ G4(k) \ ; \ v_{m2}^-=N \ G3(k) \ ; 
\label{aa}
\eea
\bea
\mbox{with} \ \ \ \ G1&=& \frac{\Gamma(1-ik_+)\Gamma(ik_-)}{\Gamma(-\frac{1}{2}ik_++\frac{1}{2}ik_-+\frac{1}{2}+
\gamma)\Gamma(-\frac{1}{2}ik_++\frac{1}{2}ik_-+\frac{1}{2}-\gamma)} \ ; \nonumber \\
G2&=& \frac{\Gamma(1-ik_+)\Gamma(-ik_-)}{\Gamma(-\frac{1}{2}ik_+-\frac{1}{2}ik_-+\frac{1}{2}-
\gamma)\Gamma(-\frac{1}{2}ik_+-\frac{1}{2}ik_-+\frac{1}{2}+\gamma)} \ ; \nonumber \\
G3&=& \frac{\Gamma(1+ik_+)\Gamma(ik_-)}{\Gamma(\frac{1}{2}ik_++\frac{1}{2}ik_-+\frac{1}{2}+
\gamma)\Gamma(\frac{1}{2}ik_++\frac{1}{2}ik_-+\frac{1}{2}-\gamma)} \ ; \nonumber \\
G4&=& \frac{\Gamma(1+ik_+)\Gamma(-ik_-)}{\Gamma(\frac{1}{2}ik_+-\frac{1}{2}ik_-+\frac{1}{2}-
\gamma)\Gamma(\frac{1}{2}ik_+-\frac{1}{2}ik_-+\frac{1}{2}+\gamma)} \ . \nonumber \\
\label{gg}
\eea
 The total wave function behaves asymptotically as,
\bea
\psi_{m}^+ (z\rightarrow+\infty)&=&A \ Ne^{ik_+z}+B \ Ne^{-ik_+z} \ ;  \nonumber \\
\psi_{m}^- (z\rightarrow -\infty)&=&N\left[\left(A \ G2 + B \ G4\right )e^{ik_-z}+\left(A \ G1 + B \ G3\right )e^{-ik_-z}\right] \ ; 
\label{m1+-}
\eea
The left and right handed reflection amplitudes which can be constructed by using the Eqs. (
\ref{rl}), (\ref{rr}) and (\ref{aa}) (or also can be calculated by (\ref{m1+-})) are written as,
\beq
r_l=\frac{G1(k)}{G2(k)} \ ; \ r_r=-\frac{G4(k)}{G2(k)} \ .
\eeq
The condition for reciprocity becomes              
\beq
\vert G1(k)\vert=\vert G4(k)\vert
\label{m1r}
\eeq
This can always be  seen directly from equation (6).

Eq. (\ref{m1r}) is only satisfied for discrete values of '$k$' i.e. only for discrete 
values of incident
particle energy (Fig. 1). Thus for this PT-symmetric non-Hermitian potential one gets 
reciprocity for certain particle energies without obtaining unitarity. Fig. 1
is showing the discrete incidence energies for which the scattering is reciprocal 
even for a PT symmetric non-Hermitian system. \\

\includegraphics[scale=1.1]{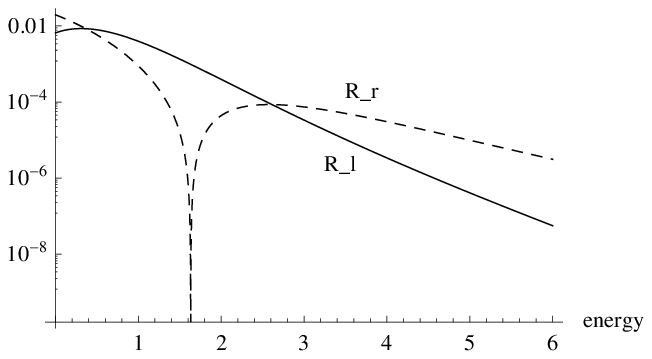} (a) \ \includegraphics[scale=1.1]{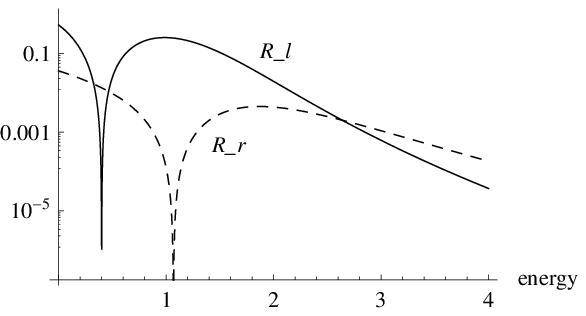} (b) \\

{\bf Fig.1:} {\it Shows different points where reciprocity is restored even for 
PT-symmetric non-Hermitian system. In 1(a) we have two energy points (in atomic unit) where $R_l=R_r$ for 
$v=1.9$ and $\mu=\pi/5$. 1 (b) shows three such energy points (with $\mu=\pi/10$) where reciprocity is restored. }  \\

\subsection{Parity violating non-Hermitian double-delta potential}
Let us consider the following double-delta potential ,
\begin{equation} 
V^{II}(x)=  \lambda  [\delta(x- \frac {a}{2})-\delta(x+ \frac {a}{2})] 
\end{equation}
where $\lambda$ and $a$ are constant parameters.
This potential becomes (i) PT-symmetric non-Hermitian for imaginary $\lambda $ and
(ii) non PT-symmetric non-Hermitian for complex $\lambda $. In either cases the non-Hermitian
potential is parity violating.
The reflection amplitudes for left and right incident particle with energy $E$ are written as,
\bea 
r_l&=&\frac {\frac{2 i \lambda }{2 k} (1-\frac{\lambda} {2k})\sin (k a)}{\left [1+(\frac{\lambda} {2k})^2
(e^{2ika}-1)\right ]} \ \ ; \nonumber \\
r_r&=&-\frac {\frac{2 i \lambda }{2 k} (1+\frac{\lambda} {2k})\sin (k a)}{\left [1+(\frac{\lambda} {2k})^2
(e^{2ika}-1)\right ]} \ \ ; \ \ \ \mbox{with} \ \ \ k=\sqrt{E} \ \ \mbox{in atomic unit}. 
\label{drr}
\eea
It is easy to see from these equations of reflection amplitudes that scattering due to this
non-Hermitian double delta potential is non-reciprocal independent of the fact that PT-symmetry
is broken or not. Further we show that reciprocity is achieved at some special situations where 
spectral singularity or reflectionlessness occurs. From Eq. \ref{drr} it is easily seen that 
at the energies $E_*$ for which $\left [\cos (2k_*a)+i\sin(2k_*a)\right ]+(\frac{k_*} 
{\lambda })^2=1$ both $r_l$ and $r_r$ are infinite. This is nothing but the condition of 
spectral singularity for the energy of the particle incident from either side. On the other 
situation reciprocity 
is restored due to the case of 
bidirectional reflectionlessness ($r_l=r_r=0$) at discrete incidence energies 
$E_{**}=\frac{n^2\pi^2}{2a^2m}$. Fig. 2
explains both the conditions where the left and right handed reflection amplitudes behave
in exactly similar way. \\

\includegraphics[scale=1.26]{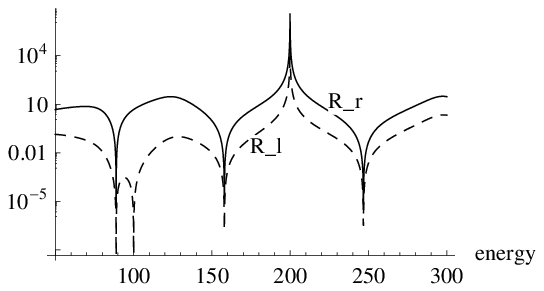} (a) \ \includegraphics[scale=1.26]{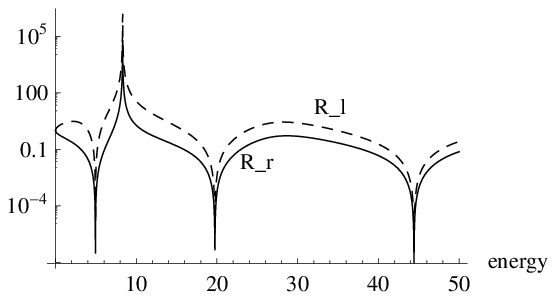} (b) \\

{\it Fig. 2: (a, b) shows the restoration of  reciprocity for parity asymmetric double delta 
non-Hermitian potential when $a=1, \lambda  =20 i$ and $a=-1, \lambda  =2.01 i - 6.1$ respectively at different discrete energies. Reciprocity and SS or 
reflectionlessness occurs at the same energy values.} \\

\subsection{Penetrating state of a complex potential and reciprocity}
The potential in Eq. \ref{realm} is complexified in PT-symmetric manner
by taking the parameter $d$ imaginary (i.e. $d\rightarrow id$) and keeping the other 
parameters real. For this case
the time independent Schroedinger equation (TISE) takes the form, 
\beq
\frac{\hbar^2}{2 m d^2}\frac{d^2\psi}{dz^2}+V^I(z)\psi =E\psi
\label{schr}
\eeq
where z is taken as $z=-iX/d-\mu$, with $X=x+i\zeta $.
Note that the differential term in this equation comes with wrong sign due to the presence 
of $d^2$ term.
However this equation can be interpreted as TISE for a upside down potential of the original 
one with energy eigenvalues (-E). 
It behaves like a potential barrier with a maxima at x=0. We have penetrating 
state solutions
if the particle has negative energy with magnitude less than $ve^{-2\mu}$. On the other hand 
it accepts 
free states when the energy of the particle is more than the barrier height.

The penetrating state solutions are given by,
\bea
\psi_1 (z) &=& Ne^{-az}(e^z+e^{-z})^{-bz} 
F\left(b+\frac{1}{2}-\sqrt{-v\cosh^2 (\mu) +1/4},\right.\nonumber \\
&& \left. b+\frac{1}{2}+\sqrt{-v\cosh^2 (\mu) 
+1/4} ;\mid a+b+1\mid;\frac{e^{-z}}{e^z+e^{-z}}\right)  \nonumber \\
\psi_2 (z) &=& Ne^{bz}(e^z+e^{-z})^{az} 
 F\left(-a+\frac{1}{2}-\sqrt{-v\cosh^2 (\mu) +1/4},\right.\nonumber \\
&&\left. -a+\frac{1}{2}+\sqrt{-v\cosh^2 (\mu) 
+1/4} ;\mid 1-a-b\mid;\frac
{e^{-z}}{e^z+e^{-z}}\right)  \nonumber \\
\label{penn}
\eea
with 
\beq
a^2+b^2=-\epsilon -v\cosh 2\mu; \ \ \ 2ab=-v\sinh2\mu ; 
\eeq
and $k_+=\sqrt{\epsilon +ve^{2\mu}}$, $k_-=\sqrt{\epsilon +ve^{-2\mu}}$
so that 
\beq
a=-1/2(ik_++ik_-); \ \ \ b=1/2(-ik_++ik_-) .
\eeq
The wave function for this penetrating state is written in a general form as,
\beq
\psi(z)=A\psi_1(z)+B\psi_2(z)
\label{ppsi}
\eeq
We calculate the left and right handed 
penetrating amplitudes from the asymptotic behavior of Eq. (\ref{ppsi}) as,
\beq
 r_l=\frac{P1}{P2} \ \ \ r_r=-\frac{P4}{P2}, 
\label{rtlp}
\eeq
where P1, P2, P3, P4 are expressed in terms of Gamma function written as,
\bea
P1&=& \frac{\Gamma(1-ik_+)\Gamma(-ik_-)}{\Gamma(-\frac{1}{2}ik_+-\frac{1}{2}ik_-+\frac{1}{2}+
\gamma')\Gamma(-\frac{1}{2}ik_+-\frac{1}{2}ik_-+\frac{1}{2}-\gamma')} \nonumber \\
P2&=& \frac{\Gamma(1-ik_+)\Gamma(ik_-)}{\Gamma(-\frac{1}{2}ik_++\frac{1}{2}ik_-+\frac{1}{2}-
\gamma')\Gamma(-\frac{1}{2}ik_++\frac{1}{2}ik_-+\frac{1}{2}+\gamma')} \nonumber \\
P3&=& \frac{\Gamma(1+ik_+)\Gamma(-ik_-)}{\Gamma(\frac{1}{2}ik_+-\frac{1}{2}ik_-+\frac{1}{2}+
\gamma')\Gamma(\frac{1}{2}ik_+-\frac{1}{2}ik_-+\frac{1}{2}-\gamma')} \nonumber \\
P4&=& \frac{\Gamma(1+ik_+)\Gamma(ik_-)}{\Gamma(\frac{1}{2}ik_++\frac{1}{2}ik_-+\frac{1}{2}-
\gamma')\Gamma(\frac{1}{2}ik_++\frac{1}{2}ik_-+\frac{1}{2}+\gamma')} \nonumber \\
\eea
where $\gamma'=\sqrt{-v\cosh^2 (\mu)+1/4} $. Here we note $r_l\not=r_r$ but $P1^*=P4; \ \ 
P2^*=P3$ (due to real $k_+$ and $k_-$ ). Thus the second condition of Eq. (\ref{ler}) is being
satisfied so that $$ \ R_l\equiv \mid r_l\mid^2=\mid r_r\mid^2\equiv R_r  \ .$$
This implies that we have reciprocity for this non-Hermitian PT-symmetric 
system for all values of energy. However as expected unitarity is violated, i.e $R+T\not=1$ 
(both for left and right handed cases) for 
this model.
We further observe no SS is present for penetrating states and the potential never becomes 
reflectionless.

\section {Conclusion} 
Scattering due to real potential is always reciprocal. Further it has been shown that even complex 
scattering obeys reciprocity if the potential is parity symmetric. This implies that scattering  
due to PT symmetric non-Hermitian systems are supposed to be non-reciprocal as PT symmetric non-Hermitian system
is essentially parity violating. In this present work we show that reciprocity is valid even for certain 
PT symmetric non-Hermitian systems. In support of our claim we consider, three explicit examples. 
The scattering states of PT-symmetric complex Morse potential is reciprocal only at certain specific 
incident energies. In another example we consider PT-symmetric as well as non PT-symmetric complex double delta potential to show 
restoration of reciprocity only at SS and reflectionless points. Most interesting result is for penetrating states
of PT symmetric non-Hermitian Morse potential which is shown to be reciprocal at all incident energy.
Unitarity of scattering S-matrix and reciprocity are two important characteristics of scattering theory.
Their validity/non validity in the case of complex scattering is extremely important.
Our results are one step forward in the investigation of reciprocity of complex scattering. No example of PT-symmetric non-Hermitian potentials is known so far
when both unitarity and reciprocity hold good. It will be worth 
finding more and more parity violating non-Hermitian systems which are reciprocal as well 
as unitary. \\

{\bf Acknowledgments}
BKM and BPM acknowledge the financial support from the Department of Science 
and Technology (DST), Gov. of India
under SERC project sanction grant No. SR/S2/HEP-0009/2012. AG acknowledges the 
Council of Scientific \& Industrial Research (CSIR), India for Senior Research Fellowship.


\begin{thebibliography}{99}

\bibitem{ben4} C. M. Bender and S. Boettcher, {\em Phys. Rev. Lett.} {\bf 80}, 5243 (1998).
\bibitem{mos} A. Mostafazadeh, {\em Int. J. Geom. Meth. Mod. Phys.} {\bf 7}, 1191(2010) and references therein.
\bibitem{benr} C.M. Bender, {\em Rep. Progr. Phys.} {\bf 70}, 947 (2007) and references therein.




\bibitem{opt1} Z. H. Musslimani, K. G. Makris, R. El-Ganainy, and D. N. Christodoulides, {\em Phys. Rev. Lett.} {\bf 100}, 030402 (2008).

\bibitem{opt2} C. E. Ruter, K. G. Makris, R. El-Ganainy, D. N. Christodoulides, M. Segev, D. Kip, {\em Nature Phys.} {\bf 6} 192, (2010); 

\bibitem{opt3}  R. El-Ganainy, K. G. Makris, D. N. Christodoulides and Z. H. Musslimani, {\em Opt. Lett.} {\bf 32}, 2632 (2007).

\bibitem{eqv1} A. Guo et al, {\em Phys. Rev. Lett.} {\bf 103}, 093902 (2009).


\bibitem{ep0} T. Kato, {\em Perturbation Theroy of Linear Operators}, {\bf Springer}, Berlin, (1966).
\bibitem{ep1} M. V. Berry, {\em Czech. J. Phys.} {\bf 54}, 1039 (2004).
\bibitem{ep2} W. D. Heiss, {\em Phys. Rep.} {\bf 242}, 443 (1994).


\bibitem{ss1} A. Mostafazadeh, {\em Phys. Rev. Lett.} {\bf 102}, 220402 (2009).

\bibitem{ss2} A. Mostafazadeh, M. Sarisaman, {\em Phys. Lett. A} {\bf 375}, 3387 (2011).

\bibitem{aop} A. Ghatak, R. D. Ray Mandal, B. P. Mandal, {\em Ann. of Phys.} {\bf 336}, 540 (2013).

\bibitem{ss3} A. Ghatak, J. A. Nathan, B. P. Mandal, and Z. Ahmed, {\em J. Phys. A: Math. 
Theor.} {\bf 45},  465305 (2012).

\bibitem{inv2}  S. Longhi, {\em J. Phys. A: Math. Theor.} {\bf 44}, 485302 (2011).

\bibitem{inv1} A. Mostafazadeh, {\em Phys. Rev. A} {\bf 87}, 012103 (2013).

\bibitem{resc} L. Deak, T. Fulop, {\em Ann. of Phys.} {\bf 327}, 1050 (2012).

\bibitem{cc0} M. Cai, O. Painter, and K. J. Vahala, {\em Phys. Rev. Lett. } {\bf 85}, 74 (2000).

\bibitem{cc1} J. R. Tischler, M. S. Bradley, and V. Bulovic, {\em Opt. Lett.} {\bf 31},
2045 (2006)

\bibitem{cc2} S. Dutta Gupta, {\em Opt. Lett.} {\bf 32}, 1483 (2007).

\bibitem{cc3} S. Balci, C. Kocabas, and A. Aydinli, {\em Opt. Lett.} {\bf 36}, 2770 (2011).

\bibitem{cc4} S. Balci, Er. Karademir, C. Kocabas, and A. Aydinli, {\em Opt. Lett.} {\bf 36}, 3041 (2011).


\bibitem{cpa00} C. F. Gmachl, {\em Nature} {\bf 467}, 37 (2010).

\bibitem{cpa01} S. Longhi, {\em Physics} {\bf 3}, 61(2010).

\bibitem{cpa011} W. Wan, Y. Chong, L. Ge, H. Noh, A. D. Stone, H. Cao, {\em Science} {\bf 331}, 889 (2011).


\bibitem{hsn} M. Hasan, A. Ghatak, B. P. Mandal, {\em Ann. of Phys. } {\bf 344 },  17 (2014).
\bibitem{ent} A. Ghatak and B. P. Mandal, {\em J. Phys. A: Math. Theor.}
{\bf 45}, 355301 (2012).

\bibitem{bmsm} B. P. Mandal and S S. Mahajan  {\em arXiv:1312.0757}, (2013).

\bibitem{ph} B. P. Mandal, B. K. Mourya, and R. K. Yadav (BHU),
{\em Phys. Lett. A} {\bf 377}, 1043 (2013).

\bibitem{bpm} B. P. Mandal, {\em Mod. Phys. Lett. A} {\bf 20}, 655(2005).

\bibitem{cal} B. P. Mandal and A. Ghatak, {\em J. Phys. A: Math. Theor.}
 {\bf 45}, 444022 (2012) .



\bibitem{cpa02} N. Liu, M. Mesch, T. Weiss, M. Hentschel, and H. Giessen, {\em Nano Lett.} {\bf 10}, 2342 (2010).

\bibitem{cpa1} H. Noh, Y. Chong, A. Douglas Stone, and Hui Cao, {\em Phys. Rev. Lett.} {\bf 108}, 6805 (2011).

\bibitem{cpa0} A. Mostafazadeh and M. Sarisaman, {\em Proc. R. Soc. A} {\bf 468}, 3224 (2012).

\bibitem{cpa2} S. Longhi, {\em Phys. Rev. A} {\bf 83}, 055804 (2011).

\bibitem{cpa3} S. Dutta-Gupta, R. Deshmukh, A. Venu Gopal, O. J. F. Martin, and S.
Dutta Gupta, {\em Opt. Lett.} {\bf 37}, 4452 (2012).

\bibitem{cpa4} N. Liu, M. Mesch, T. Weiss, M. Hentschel, and H. Giessen, {\em Nano Lett.} {\bf 10} 2342 (2010).

\bibitem{reci1} Z. Ahmed, {\em Phys. Lett. A} {\bf 377}, 957 (2013).
\bibitem{reci2} A Mostafazadeh, {\em arXiv:1405.4212}, (2014).

\bibitem{ram} Ram Narayan Deb, Avinash Khare, Binayak Dutta Roy, {\em Phys. Lett. A} {\bf 307}  215 (2003).


\end{thebibliography}
\end{document}